\documentstyle[pre,aps,multicol,epsfig]{revtex}
\begin{document}

\title{Stable two-dimensional dispersion-managed soliton}

\author{Fatkhulla Kh. Abdullaev$^{1,2}$, Bakhtiyor B. Baizakov$^{2,3}$,
        and Mario Salerno$^3$}

\address{$^1$ Instituto de Fisica Teorica, UNESP, Rua Pamplona 145, \\
              01405-900, Sao Paulo, Brasil \\
         $^2$ Physical-Technical Institute, Uzbek Academy of Sciences, \\
              2-b, Mavlyanov str., 700084, Tashkent, Uzbekistan \\
         $^3$ Dipartimento di Fisica "E.R. Caianiello" and Istituto
              Nazionale di Fisica  \\
              della Materia (INFM), Universit\'a di Salerno, \\
              I-84081 Bareness (SEA), Italy }

\maketitle

\begin{abstract}

The existence of a dispersion-managed soliton in two-dimensional
nonlinear Schr\"odinger equation with periodically varying
dispersion has been explored. The averaged equations for the
soliton width and chirp are obtained which successfully describe
the long time evolution of the soliton. The slow dynamics of the
soliton around the fixed points for the width and chirp are
investigated and the corresponding frequencies are calculated.
Analytical predictions are confirmed by direct PDE and ODE
simulations. Application to a Bose-Einstein condensate in optical
lattice is discussed. The existence of a dispersion-managed
matter-wave soliton in such system is shown.

\end{abstract}


\begin{multicols}{2}

\section{Introduction}

Nonlinear wave propagation in media with periodically varying
dispersion is attracting a huge interest over the recent years. A
prominent example is a dispersion-managed (DM) optical soliton,
which is considered to become the major concept in future
soliton-based communication systems. It was shown theoretically
and experimentally that the strong DM regime provides the
undisturbed propagation of pulses over very long distances. DM
solitons are robust to the Gordon-Haus timing jitter, which makes
them favorable against the standard solitons
\cite{Doran,Gabitov1}.

Mathematically this type of problems are described by the one
dimensional (1D) nonlinear Schr\"odinger (NLS) equation with
periodic dispersion - a nonlinear analogue of the Mathieu
equation. The corresponding linear equation exhibits a rich
variety of stability and instability zones for the parameters. The
existence of a DM soliton is one of the nontrivial consequences of
the stable diagram for the periodic NLS equation.

Although well studied in the 1D case, the two and three
dimensional extensions of this problem are far less explored. The
major difference here is that, contrary to the 1D case, the NLS
equation in two and three dimensions is unstable against collapse.
In particular, for the two dimensional (2D) case the collapse
occurs if the initial power exceeds some critical value, i.e. if
$E > E_{cr}$. Recently it has been demonstrated that the {\it
nonlinearity} management can prevent the collapse of solitons in
2D Kerr type optical media \cite{Berg,Towers}, as well as in 2D
Bose-Einstein condensates \cite{Abd2,Sai}. From these one can
reasonably expect that the dispersion-management can play
balancing role also in the 2D case, and the stable 2D DM soliton
can exist. Such a possibility has recently been considered in
Ref.\cite{Zhar} by construction of the ground state for the
periodic 2D NLS equation based on the averaged variational
principle and the techniques of integral inequalities - i.e. the
proof of the existence theorem for DM soliton was presented.
Analytical and numerical treatment of the problem, however, has
not been addressed so far.

The purpose of this Communication is to derive  analytical
expressions for the parameters of a 2D dispersion-managed
soliton and to study the conditions for their stability. To this
regard, we use a time-dependent variational approach (VA) to
derive a set of ODEs for the soliton parameters. The stability of
the DM soliton is then inferred from the stability of fixed points
of the VA equations.

The field dynamics is governed by the following 2D NLS equation
\begin{equation} \label{nlse2d}
  iu_{t} + d(t)\Delta u + |u|^{2} u = 0,
\end{equation}
where $d(t) = d_0 + d_1(t)$ represents a time periodically varying
dispersion coefficient. In the strong DM regime it is assumed that
$d(t) \sim (1/\epsilon)d(t/\epsilon), \epsilon
\ll 1$ and the dispersion averaged over the period is $<d(t)> =
d_{0}$ (in this case  $d_{0}> 0$ corresponds to a negative
dispersion and $d_{0} < 0$ to a positive one).

The equation (\ref{nlse2d}) can be associated with two main physical
problems: (i) beam propagation in 2D waveguide arrays with
periodically variable coupling between waveguides
\cite{peschel,darmanyan};  (ii) nonlinear matter-waves of
Bose-Einstein condensates in 2D optical lattices.

In case (i) the model equations for a 2D nonlinear fiber array are
given by \cite{Aceves}
\begin{equation}
  i\psi_{n,z} +\kappa(z)\Delta_{2}\psi_{n} + \omega''\psi_{n,tt} + \chi
  |\psi_{n}|^2 \psi_{n}=0,
\end{equation}
where $\psi_{n}$ is the envelope of electric field in the n-th fiber,
$\Delta_{2}$ is the finite second difference for 2D, $\kappa(z)$
is the variable along $z$ coupling coefficient
\cite{peschel,darmanyan}, $\omega''$ is the group-velocity
dispersion, $\chi$ is the coefficient of nonlinearity. For long
wavelength pulses the group-velocity dispersion $\omega''$ can be
neglected. Introducing the dimensionless variables $\kappa z = t,
\psi_{n} = \sqrt{2\kappa/\chi}u_{n}$ and considering the field
distribution to be broad in the transverse direction ($>7$ sites),
one arrives to Eq. (\ref{nlse2d}) with time and space interchanged
and with $d(t)$ describing a varying diffraction along the
longitudinal direction. Notice that although the intrinsic
discreteness of the array may arrest the collapse of a 2D NLS
wave, it does not necessarily stabilize the pulse against decay. In
the following we show that this can be done employing dispersion
(diffraction) management by means of which a stable 2D soliton can
be created before the strong shrinking of the wave occurs.

A similar situation arises in case (ii) for a Bose-Einstein
condensates (BEC) confined in a 2D optical lattice. In this case
dynamics of the condensate is governed by the Gross-Pitaevskii
(GP) equation
\begin{equation}
  i\hbar\Psi_{t} =-\frac{\hbar^2}{2m}\Delta\Psi + V(x,y;t)\Psi + g_{2D}|
\Psi|^{2}\Psi,
\end{equation}
where $g_{2D} = g_{3D}/\sqrt{2}\pi a_{z}), g_{3D} = 4\pi \hbar^2
a_{s}/m, a_{z} =(\hbar/m\omega_{z})^{1/2}$ and  with $V(x,y) =
V_{0}(t)(\cos^{2}(k_{0}x) + \cos^{2}(k_{0}y))$ denoting an optical
lattice with the amplitude periodically varying in time.
Spatiotemporal wave collapse in the framework of a similar equation
(when the potential is periodic in one direction $V(x,y) = V_0 \cos(k_0 x)$)
was considered in \cite{kivshar94}, where analytical expression for the
upward shift of collapse criterion was derived for potentials rapidly
oscillating in space (large $k_0$).
By adopting an effective mass description one can show that the 2D GP
equation \cite{ks} can be reduced to the DM NLS equation
(\ref{nlse2d}). The effectiveness of DM applied to quasi-1D atomic
matter-waves was experimentally demonstrated in
Ref.\cite{eiermann}.

For analytical considerations it is convenient to
refer to the axially symmetric case for which $\Delta
=\partial^{2}/\partial r^{2} + (1/r)(\partial/\partial r)$, and
apply the harmonic modulation for dispersion-management: $d(t) =
d_{0} + d_{1}\sin(\Omega t)$. We remark that although in the
present Communication we do not consider the case of two-step
dispersion-management: $d(t) = d_{+}, \mbox{ if } \ t_+ + nt_p
> t > n t_p, \mbox { and } d(t) = d_{-}, \mbox{ if } (n+1) t_p > t >
nt_p+t_+$, where $ t_p=t_+ + t_- , \ \ n=0, 1, 2...$ , this approach
can also be effectively used for the creation of stable 2D DM
solitons.

Our analysis of the pulse dynamics under dispersion-management is
based on the variational approach \cite{Gabitov1,Malomed},
according to which a space averaged Lagrangian $\bar{L} = \int L
d{\bar r}$ is constructed starting from a suitable ansatz for the
soliton profile. In the following we shall calculate $\bar{L}$ by
using the following Gaussian ansatz
\begin{equation} \label{ansatz}
 u(r,t) = A(t)\exp(-\frac{r^{2}}{2a^{2}} + i\frac{b(t)r^{2}}{2} + i\phi(t)),
\end{equation}
where $A, a, b, \phi$ denote the amplitude, width, chirp and
linear phase of the soliton, respectively. The equations for the
soliton parameters are then derived from the Euler-Lagrange
equations for $\bar{L}$ as
\begin{equation} \label{vareq}
  a_{t} =2d(t)\beta, \quad  \beta_{t} = \frac{2d(t)-E}{a^3},
\end{equation}
where $\beta = ab$, and
$E = \int_0^{\infty} |u|^2 r dr$ is the energy.

\section{System of averaged variational equations}

Let us consider the evolution of a pulse (a beam or a soliton
matter-wave, depending on the physical system in consideration)
using the division on the fast and slow time scales
\cite{Abd1,Tur1,Garnier}. The width and chirp of the pulse are
then represented as $a(t) = \bar{a} + a_{1}, \ \beta(t) =
\bar{\beta} + \beta_{1}$, where $\bar{a},\bar{\beta}$ are slowly
varying functions on the scale $1/\epsilon$ and $a_{1}, \beta_{1}$
are rapidly varying functions. The solutions for $a_{1},
\beta_{1}$ are
\begin{eqnarray}
  a_{1} = -\frac{4d_{0}d_{1}}{\bar{a}^{3}(\omega_{0}^{2} +\Omega^{2})}
  \sin(\Omega t)
  - \frac{2\Omega d_{1}\bar{\beta}}{\omega_{0}^2 + \Omega^{2}}\cos(\Omega t),\\
  \beta_{1} = \frac{6\sigma d_{1}\bar{\beta}}{\bar{a}^{4}(\omega_{0}^2 +\Omega^{2})}
  \sin(\Omega t) - \frac{2d_{1}\Omega}{\bar{a}^{3}(\omega_{0}^2 + \Omega^{2})}
  \cos(\Omega t),
\end{eqnarray}
where $\omega_{0}^2 = -6\sigma/\bar{a}^{4}$, $\sigma = 2d_{0} - E$.
Note that $\sigma <0$ for over-critical energy for collapse
$E > E_{cr} = 2$, at $d_0 = 1$ given by the VA. The exact value,
corresponding to the so called "Townes soliton" is $E_{cr} =
1.862$ \cite{berge}. Considering the limit of high frequencies
$\Omega^2 \gg \omega_{0}^{2} \sim 1 $, for the averaged parameters
of the system we finally get
\begin{eqnarray}\label{av1}
  \bar{a}_{t} = 2\bar{\beta} (d_{0} + \frac{3 d_{1}^{2}\sigma}
  {\Omega^{2}\bar{a}^4}),
  \\ \label{av2}
  \bar{\beta}_{t} = \frac{\sigma}{\bar{a}^3} + \frac{12d_{1}^{2} d_{0}}
  {\Omega^{2}\bar{a}^7} + \frac{12\sigma d_1^2 \bar{\beta}^2}{\Omega^2 \bar{a}^5}.
\end{eqnarray}
This system has the Hamiltonian structure with the Hamiltonian given
by
\begin{equation}\label{Ham}
  H = \frac{\sigma}{2{\bar{a}}^2} + \frac{2\Lambda^2 d_{0}}{{\bar{a}}^6} +
  \bar{\beta^2} (d_{0} + \frac{3\Lambda^{2}\sigma}{\bar{a}^4}), \quad
  \Lambda = \frac{d_1}{\Omega},
\end{equation}
from which the equations of motion follows as $\bar{a}_{t} =
\partial H/\partial\bar{\beta}, \ \bar{\beta}_{t} = -\partial
H/\partial \bar{a}.$ From this Hamiltonian one can also see that
the mechanism for collapse suppression originates from the
repulsive potential near the small values of width $\sim
1/\bar{a}^6$, which counteracts to the attractive force induced by
the nonlinearity $\sim 1/\bar{a}^2$. The exact balance between
these forces gives rise to a stable state. This state is
oscillatory with the frequency which will be defined later. The
stabilization mechanism of a 2D NLSE soliton is similar to that of
the inverted pendulum with oscillating pivot point \cite{Landau}.
We should note that the averaged dynamics is not potential - a
velocity dependent term appears in the interaction potential (see
4th term in (\ref{Ham})). Although this term doesn't contribute to
the fixed point, it is important for the description of
oscillatory dynamics of 2D DM solitons.

The system (\ref{av1}),(\ref{av2}) has the fixed points
\begin{equation} \label{fixp}
  \bar{\beta} =0, \ \bar{a}_{c} = (-\frac{12 d_{0}\Lambda^2}{\sigma})^{1/4}.
\end{equation}
\begin{figure}[ht]
\centerline{\includegraphics[width=8cm,height=6cm,clip]{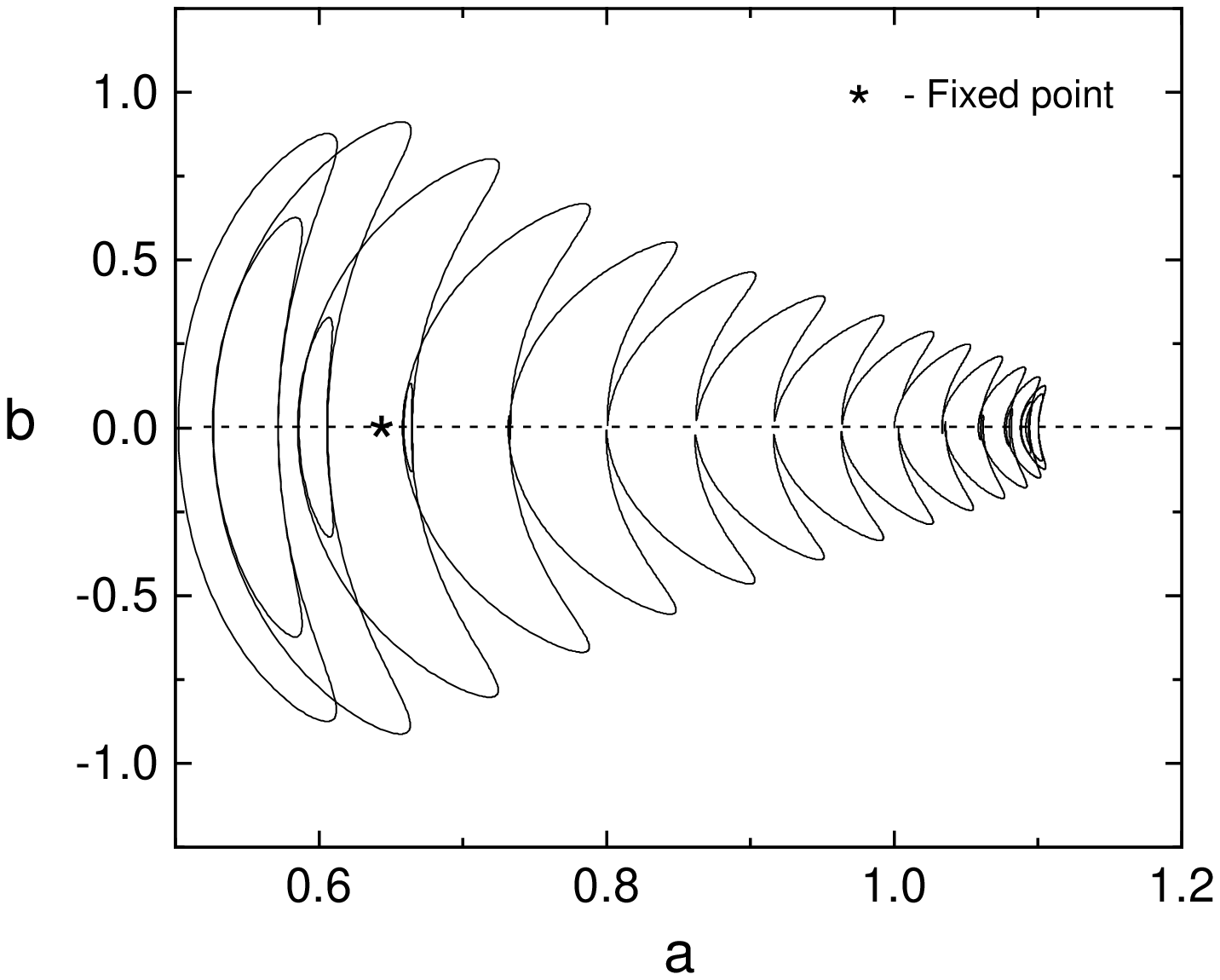}}
\vspace*{0.5cm} \caption{Phase portrait of the variational
equations (\ref{vareq})
         with parameters $d_0 = 1, \ d_1 = 3.5, \ \Omega = 50, \
         E = N/2\pi = 2.3034$.}
\label{fig1}
\end{figure}
Note that $\Lambda$ is proportional to the strength of the
dispersion map $D = 2 \pi d_1/\Omega$, therefore $\bar{a}_{c} \sim
\sqrt{D}$ in analogy with the estimate for a DM soliton in 1D
case. There exists one solution with a stationary width for the
anomalous residual dispersion $d_{0} > 0, E > 2d_{0}$. This is
confirmed by the phase portrait (Fig.\ref{fig1}) of the
variational system (\ref{vareq}).

Let us analyze the stability of fixed points for the anomalous
residual dispersion $d_{0} > 0$. We assume $a = a_{c} + \epsilon
a_{1}, \beta = \epsilon\beta_{1}$. Substituting into
Eq.(\ref{av1}) and Eq.(\ref{av2}), and collecting terms of order
$\epsilon$ we find
\begin{eqnarray} \label{sl}
  a_{1,t} = (2d_{0} + \frac{6\Lambda^2 \sigma}{a^4})\beta_{1} =
            M \beta_{1}, \\
  \beta_{1,t} = -(\frac{3\sigma}{a^4} + \frac{84\Lambda^2 d_{0}}{a^8})
  a_{1} =-S a_{1}.
\end{eqnarray}
The oscillations of the width and chirp near the fixed points
are stable if $M S > 0$, which is always satisfied for
$d_{0} > 0, E > 2d_{0}$.
The frequency of secondary slow oscillations of a 2D DM soliton is
proportional to $\sqrt{M S}$.

\section{Numerical simulations}

To avoid the singularity at $r=0$ we consider the problem in
Cartesian coordinates $\Delta = \partial^2_x + \partial^2_y$, and
$r^2=x^2+y^2$. Then numerical simulations can be performed by
two-dimensional fast Fourier transform \cite{numrecipes}. The
results are produced using a 2D grid of 256 x 256 points over the
domain $x,y \in [-6.4, 6.4]$ and the time step $\delta t = 0.001$.
To prevent the back-action of a small amount of linear waves,
resulted from the periodic perturbation, the absorption on the
domain boundaries is employed, which also imitates the infinite
domain condition. The dispersion map was supposed to have
parameters $d_0 = 1, \ d_1 = 3.5, \ \Omega = 50$.
\begin{figure}[ht]
\centerline{\includegraphics[width=8cm,height=5.5cm,clip]{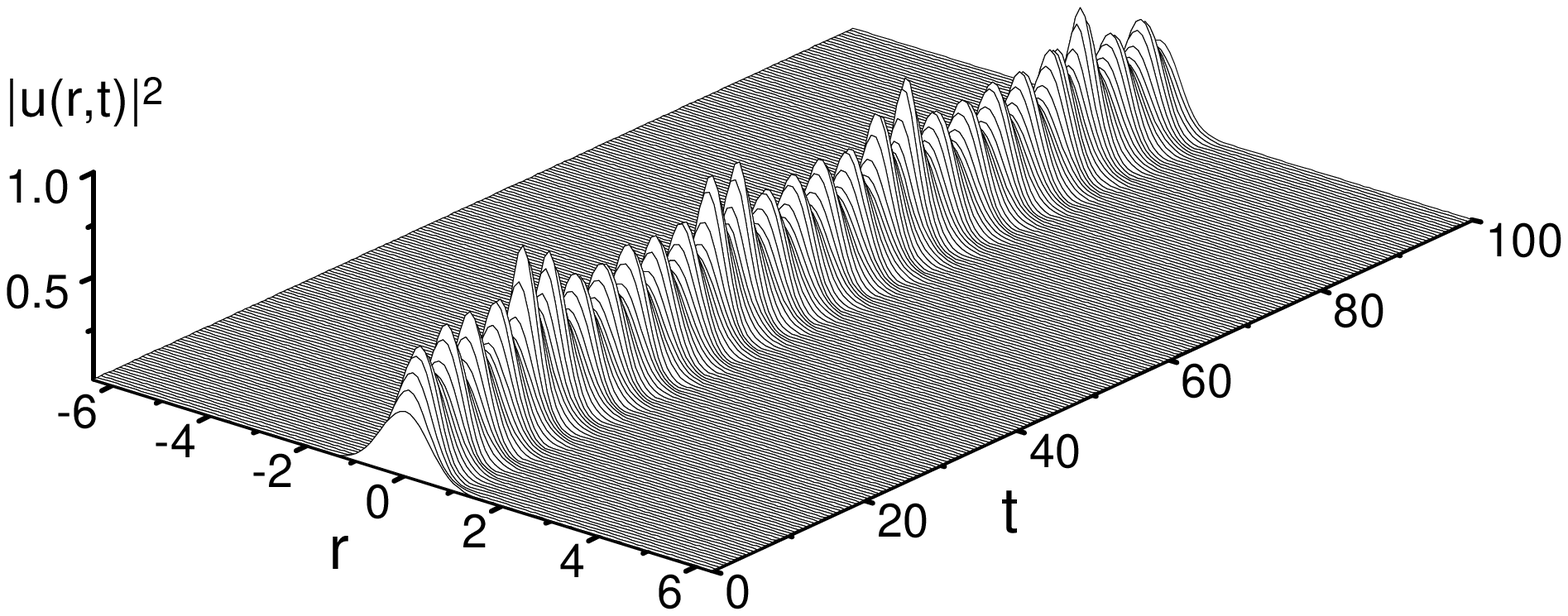}}
\vspace*{0.5cm} \caption{Evolution of a 2D DM soliton according to
         numerical solution of the equation (\ref{nlse2d}).
         The wave function is normalized to $N = 2 \pi E_0 $ with
         $E_0 = 2.3034$, and the dispersion map is
         $d_0 = 1, \ d_1 = 3.5, \ \Omega = 50.$ }
\label{fig2}
\end{figure}
This choice of parameters corresponds to moderate
dispersion-management ($D \simeq 0.45$). The axial section profile
of the wave function $|u(r,t)|^2$ as obtained by direct numerical
solution of the PDE (\ref{nlse2d}) is presented in Fig.\ref{fig2}.
As can be seen, rather stable quasi-periodic dynamics is realized
for a selected parameter settings. Note that would the periodic
modulation of the dispersion had not been applied, the initial
waveform would have collapsed within $t \sim 3$. The
dispersion-management stabilizes the pulse against the collapse or
decay, providing undisturbed propagation over very long distances.
The agreement between the predictions of the variational equations
(\ref{vareq}) for the width of a 2D DM soliton and the
corresponding result from the full PDE simulations is reported in
Fig.\ref{fig3}. As can be observed from this figure, the width of
a 2D DM soliton performs quasi-periodic motion with the average
width of $\bar a \simeq 0.8$ according to variational equations,
while the PDE simulation yields $\bar a \simeq 0.7$. The fixed
point for the above set of parameter values, according to
eq.(\ref{fixp}) is $\bar a_c = 0.6635$ (see Fig.\ref{fig1}). The
frequencies of slow dynamics given by the VA equations and PDE are
also in well agreement (Fig.\ref{fig3}). The estimate for the
frequency of slow oscillations from Eq.(\ref{sl}) yields $\omega_a
= \sqrt{M S} = 3.5$, therefore the period is $T_a = 1.9$. The
direct gauge from the Fig.\ref{fig3} shows that $T_a \simeq 2.2$,
in reasonable agreement with the above VA estimate.

For Bose-Einstein condensates in a 2D optical lattice the dispersion
coefficient can be expressed as $d(t)=m/m^{\ast}(t)$
in the effective mass formalism \cite{ks}. The effective mass
$m^{\ast}$ substantially differs from the true mass $m$ (becoming even
negative) and can be varied by changing the parameters of the
periodic potential, or inducing the transitions between energy
bands.
\begin{figure}[ht]
\centerline{\includegraphics[width=7cm,height=5.cm,clip]{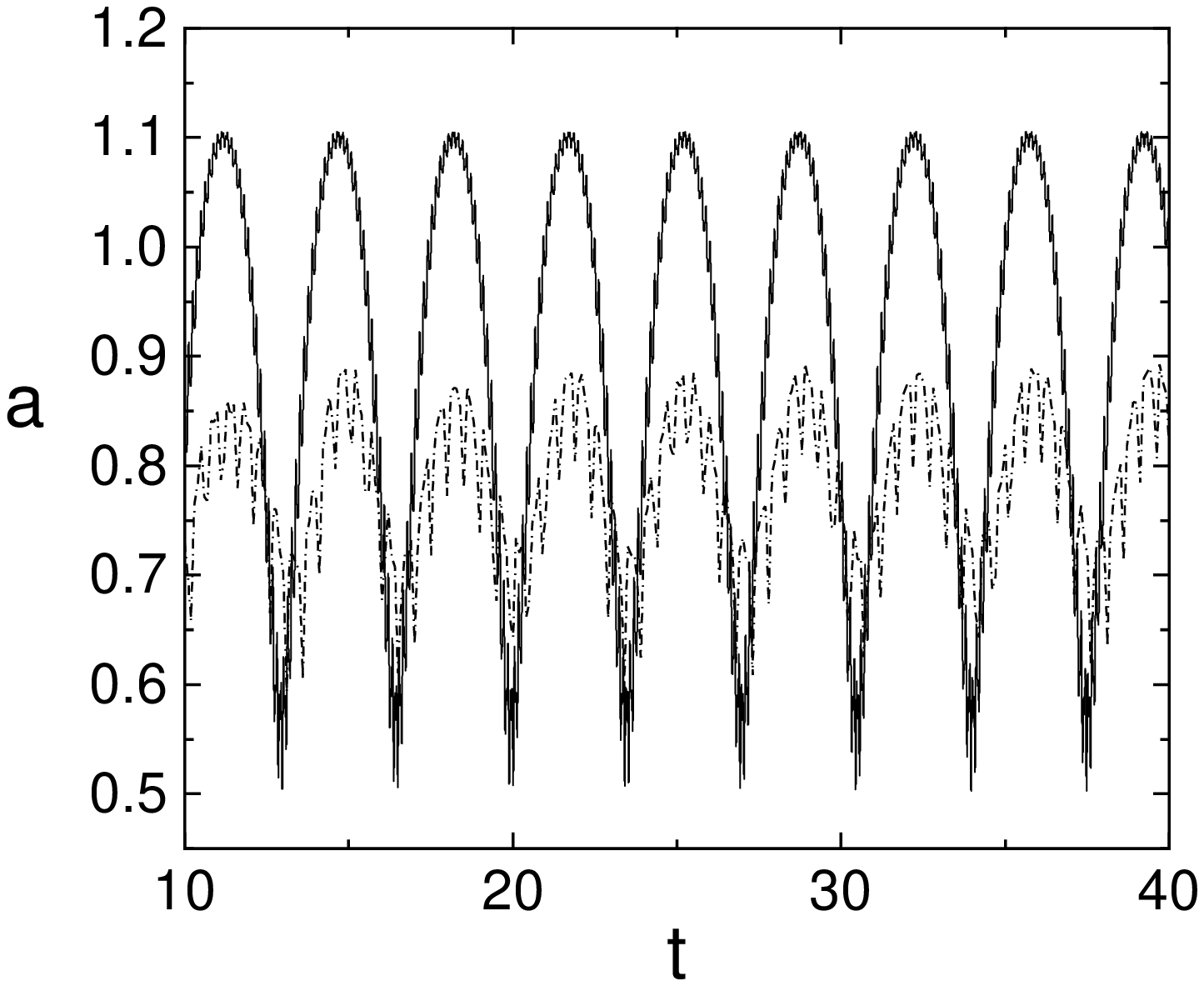}}
\vspace*{0.5cm} \caption{Stable quasi-periodic dynamics of the
width of a 2D DM soliton.
         Solid line - variational equations (\ref{vareq}) solved
         for $E = N/2\pi = 2.3034$, and the initial conditions
         $a(0) = 1, \ \beta(0) = 0$. Dashed line - full PDE
         simulations of the equation (\ref{nlse2d}). }
\label{fig3}
\end{figure}
For example, transitions between the 1st and 2nd bands (at the
band edges) in the optical lattice of strength $V_0 = 2.4 E_{rec}$
( where $E_{rec} = \hbar^2 k_0^2/2m$ is the recoil energy, $k_0 =
2\pi/\lambda_0$, $\lambda_0$ is the laser wavelength), leads to
variation of the dispersion coefficient in the range $d(t)= ( -2.5
\div 4.5 )$ as considered above.

\section{Conclusion}

In conclusion, we have demonstrated the possibility to stabilize
the 2D soliton with over-critical energy $(E > E_{cr})$ by
applying the dispersion-management. The developed theory based on
the variational approximation successfully describes the long term
evolution of a 2D DM soliton, which is confirmed by direct PDE
simulations. We discussed the possible experimental realization of
a stable 2D dispersion-managed soliton in Bose-Einsten condensates
confined to optical lattices.

\section*{Acknowledgements}

The work of F.K.A. is partially supported by FAPESP and the Uzb.AS
(Award 15-02). B.B.B. and M.S. acknowledge partial financial support
from the MIUR, through the inter-university project PRIN-2000,
and from the European grant LOCNET no. HPRN-CT-1999-00163.

\end{multicols}

\end{document}